\documentclass[prb,aps,twocolumn,superscriptaddress,floatfix,citeautoscript,showpacs]{revtex4}
\usepackage{graphicx,rotating,subfigure,amsmath,amsfonts,amssymb,delarray}
\usepackage{color}
\usepackage{dsfont}
\usepackage[T1]{fontenc}
\usepackage{bbold}

\newcommand{\e}{\text{e}}

\def\12{\frac{1}{2}}

\newcommand{\be}{\begin{equation}}
\newcommand{\ee}{\end{equation}}
\newcommand{\bea}{\begin{eqnarray}}
\newcommand{\eea}{\end{eqnarray}}

\begin{document}
\bibliographystyle{apsrev_cm}

\title{Many-body localization in infinite chains}

\author{T. Enss}
\affiliation{Institute for Theoretical Physics, University of
  Heidelberg, 69120 Heidelberg, Germany}
\author{F. Andraschko}
\author{J. Sirker}
\affiliation{Department of Physics and
  Astronomy, University of Manitoba, Winnipeg R3T 2N2, Canada}

\date{\today}

\begin{abstract}
  We investigate the phase transition between an ergodic and a
  many-body localized phase in infinite anisotropic spin-$1/2$
  Heisenberg chains with binary disorder. Starting from the N\'eel
  state, we analyze the decay of antiferromagnetic order $m_s(t)$ and
  the growth of entanglement entropy $S_{\textrm{ent}}(t)$ during
  unitary time evolution. Near the phase transition we find that
  $m_s(t)$ decays exponentially to its asymptotic value
  $m_s(\infty)\neq 0$ in the localized phase while the data are
  consistent with a power-law decay at long times in the ergodic
  phase. In the localized phase, $m_s(\infty)$ shows an exponential
  sensitivity on disorder with a critical exponent $\nu\sim 0.9$. The
  entanglement entropy in the ergodic phase grows sub-ballistically,
  $S_{\textrm{ent}}(t)\sim t^\alpha$, $\alpha\leq 1$, with $\alpha$
  varying continuously as a function of disorder.  Exact
  diagonalizations for small systems, on the other hand, do not show a
  clear scaling with system size and attempts to determine the phase
  boundary from these data seem to overestimate the extent of the
  ergodic phase.
\end{abstract} 


\maketitle 

\section{Introduction}
It is by now well established that disorder can drive closed
one-dimensional quantum many-body systems into a many-body localized
(MBL) phase \cite{Imbrie2016, Review1, Review2}. In such a phase the
system fails to act as a bath for its own subsystems and
thermalization does not occur. Instead, memory of the initial
conditions is retained. The `drosophila' to study properties of the
MBL phase is the spin-$1/2$ Heisenberg chain
\begin{equation}
\label{Ham}
H=J\sum_{i=1}^L \left(s^x_is^x_{i+1}+s^y_is^y_{i+1}+\Delta s^z_is^z_{i+1} +D_i s^z_i  \right)
\end{equation}
with $\Delta=1$ and $D_i\in [-D,D]$ a random variable drawn from a
uniform box distribution with disorder strength $D$. Here $L$ is the
length of the system and $s^\alpha_i$ is the $\alpha$ component of the
spin operator acting at site $i$. Studies of this model have been
based mainly on exact diagonalization (ED) for small systems
\cite{PalHuse, Luitz1, Luitz2, BarLev, Agarwal}. These numerical results
have then been used to determine a critical point $D_c$ between the
ergodic and MBL phase by showing, for instance, that the level
statistics changes from a Wigner-Dyson distribution at small but
nonzero $D$ to a Poisson distribution at $D>D_c$ (MBL) with
$D_c\approx 3.5$. Furthermore, deep in the MBL phase the entanglement
entropy is shown to increase logarithmically during unitary time
evolution \cite{BardarsonPollmann}, confirming results from an earlier
density matrix renormalization group study \cite{ZnidaricProsen}.

\begin{figure}
 \includegraphics*[width=1.0\columnwidth]{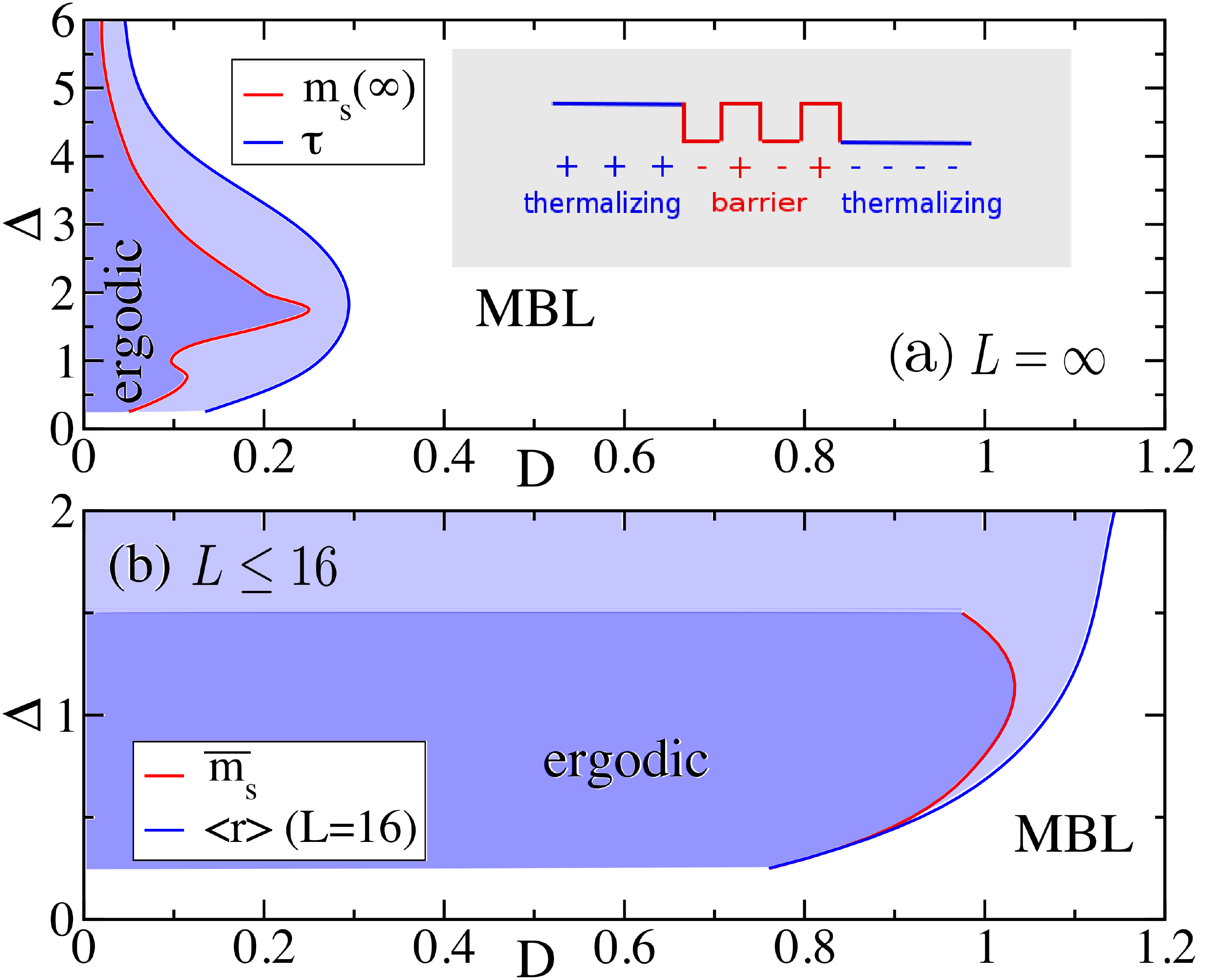} \caption{Spin chain
 \eqref{Ham} with binary disorder: (a) Phase boundary for the infinite
 chain obtained from the order parameter $m_s(\infty)$. The relaxation
 time $\tau$ changes substantially near the transition line, see
 Sec.~\ref{SecII}. Inset: Thermalizing clusters (equal field $D$) are
 separated by barriers (staggered field $\pm D$). (b) ED phase
 boundary from finite-size extrapolation $\lim_{L\to\infty}
 \overline{m_s}(L)$ and the line where the average gap in the energy
 spectrum for $L=16$ crosses the intermediate value $\langle
 r\rangle=0.4575$ between GOE and Poisson, see Sec.~\ref{SecIV}.}
\label{Fig_phase_diag}
\end{figure} 
ED studies of small systems are, however, ill-suited to address the
properties of weakly disordered systems as well as the phase
transition itself because in both cases the relevant length scale
$\xi$ will be much larger than the achievable system sizes $L$. This
creates, in particular, a significant obstacle in understanding this
novel type of dynamical phase transition where the entanglement
entropy changes from volume law (ergodic) to area law (MBL), making it
distinct from regular thermal transitions or ground state critical
points. Two approaches have so far been used to tackle this problem:
On the one hand, it has been tried to investigate the critical regime
based on extrapolations from ED data to larger systems \cite{Luitz1,
BarLev, Agarwal, Luitz2}. Assuming that the transition is described by
a single diverging length scale $\xi\sim |D-D_c|^{-\nu}$, the obtained
results are mostly consistent with a critical exponent $\nu\sim1$.
This, however, would violate a Harris-type bound which demands
$\nu>2/d$ in $d$ dimensions in order for the transition to be stable
\cite{NandkishorePotter, ChandranLaumann}. A second recent approach is
based on a real-space renormalization group (RG) applied to effective
minimal models assuming that only two energy scales exist
\cite{VoskHuse, PotterVasseur}. The length scale $\xi$ is then found
to diverge with an exponent $\nu\approx 3-3.5$ consistent with the
Harris bound. However, it is important to stress that the RG
approaches are not based on microscopic models and contradict the
results from previous ED studies.

In this work we shed new light on this controversial point by studying
a disordered interacting quantum chain directly in the thermodynamic
limit (TDL). In this way we avoid the fundamental obstacle $\xi\gg L$
one faces in ED studies of the phase transition. In the following we
focus on the anisotropic Heisenberg chain, Eq.~\eqref{Ham}, with
binary disorder $D_i=\pm D$ instead of the more commonly used box
disorder. This naturally arises as an effective model for a bosonic
system with a mobile and an immobile species in the limit of strong
onsite Hubbard interactions and also exhibits a transition from an
ergodic to an MBL phase.\cite{AndraschkoEnssSirker, TangRigol} As in
the noninteracting case\cite{ZhangSirker}, one expects that the chosen
disorder distribution leads to quantitative changes while the
qualitative features, in particular the properties of the transition,
are universal. The goals of this work are to establish the phase
diagram of the model \eqref{Ham} with binary disorder as a function of
disorder strength $D$ {\it and} anisotropy $\Delta$ (see
Fig.~\ref{Fig_phase_diag}) and to study the ergodic-MBL phase
transition directly in the TDL. In order to obtain an exact disorder
average in a single simulation, we introduce an ancilla spin-$1/2$,
$s^z_{i,\textrm{anc}}$ at each site and replace
$D_is^z_i \to 2Ds^z_is^z_{i,\textrm{anc}}$. The state of
$s^z_{i,\textrm{anc}}=\pm 1/2$ then determines the local binary
disorder $D_i=\pm D$ \cite{ParedesVerstraete,
  AndraschkoEnssSirker}. We consider the unitary time evolution
starting from an initial product state
$|\Psi(0)\rangle\otimes|\textrm{dis}\rangle$ in the Hilbert space of
spins and ancillas, where
$|\textrm{dis}\rangle=\bigotimes_j\left(\lvert\uparrow\rangle_{j,\textrm{anc}}+\lvert\downarrow\rangle_{j,\textrm{anc}}\right)/\sqrt{2}$
represents a superposition of all possible disorder
configurations. Following recent experiments \cite{SchneiderBloch1,
  SmithMonroe, SchneiderBloch2} we prepare the spins in the N\'eel
state
$\lvert\Psi(0)\rangle
=\lvert\uparrow\downarrow\uparrow\downarrow\cdots\rangle$
($|1010\cdots\rangle$ in the equivalent fermionic picture). We then
study the exactly disorder averaged decay of the antiferromagnetic
order
\begin{equation}
\label{def_ms}
m_s(t) = \langle\textrm{dis}|\langle\Psi(t)|\left(\hat{m}_s\otimes\mathbb{1}\right)|\Psi(t)\rangle|\textrm{dis}\rangle
\end{equation}
where $\hat{m_s}=L^{-1}\sum_j(-1)^j s^z_j$ measures the staggered
magnetization (imbalance) of the physical spins and the identity
operator $\mathbb{1}$ acts on the ancillas. The time evolved state
is defined by $|\Psi(t)\rangle|\textrm{dis}\rangle
=\exp(-iHt)|\Psi(0)\rangle|\textrm{dis}\rangle$ where $H$ includes the
coupling between spins and ancillas $2Ds^z_is^z_{i,\textrm{anc}}$. In
addition, we also study the growth of the disorder averaged
entanglement entropy
$S_{\textrm{ent}}=-\mbox{Tr}\rho_{\textrm{red}}\ln\rho_{\textrm{red}}$. Here
$\rho_{\textrm{red}}$ denotes the reduced density matrix of half of
the infinite chain consisting of spins and ancillas. Because the
density matrix includes the ancillas, the entanglement entropy is
quantitatively not the same as for a semi-infinite chain consisting of
spins only. The ancillas are, however, completely static so that the
entanglement entropies with and without the ancillas show the same
scaling with time.  We simulate the translationally invariant system
of spins and ancillas using the light cone renormalization group
(LCRG), a variant of the density matrix renormalization group which
yields results directly in the TDL \cite{EnssSirker,
AndraschkoEnssSirker}.

We choose the LCRG bond dimension such that the truncation error
always remains smaller than $10^{-11}$. By comparing with results
obtained keeping the truncation error smaller than $10^{-8}$ we make
sure that our results are numerically exact for the times shown. This
requires bond dimensions of up to $20000$ states. The scales we are
reaching in these simulations are unprecedented: at small disorder in
the ergodic phase correlations spread approximately ballistically
as $vt$ where $v$ is the maximal velocity of excitations in the
lattice.  The maximal times in our simulations therefore test the system at
length scales of at least $L\sim 2vt\sim 100$. While we cannot exclude
the possibility that the scaling of the quantities we study changes
qualitatively at even larger scales, our data represent substantial
progress compared to ED studies which are limited to scales of $L\sim
20$.

\begin{figure}
 \includegraphics*[width=1.0\columnwidth]{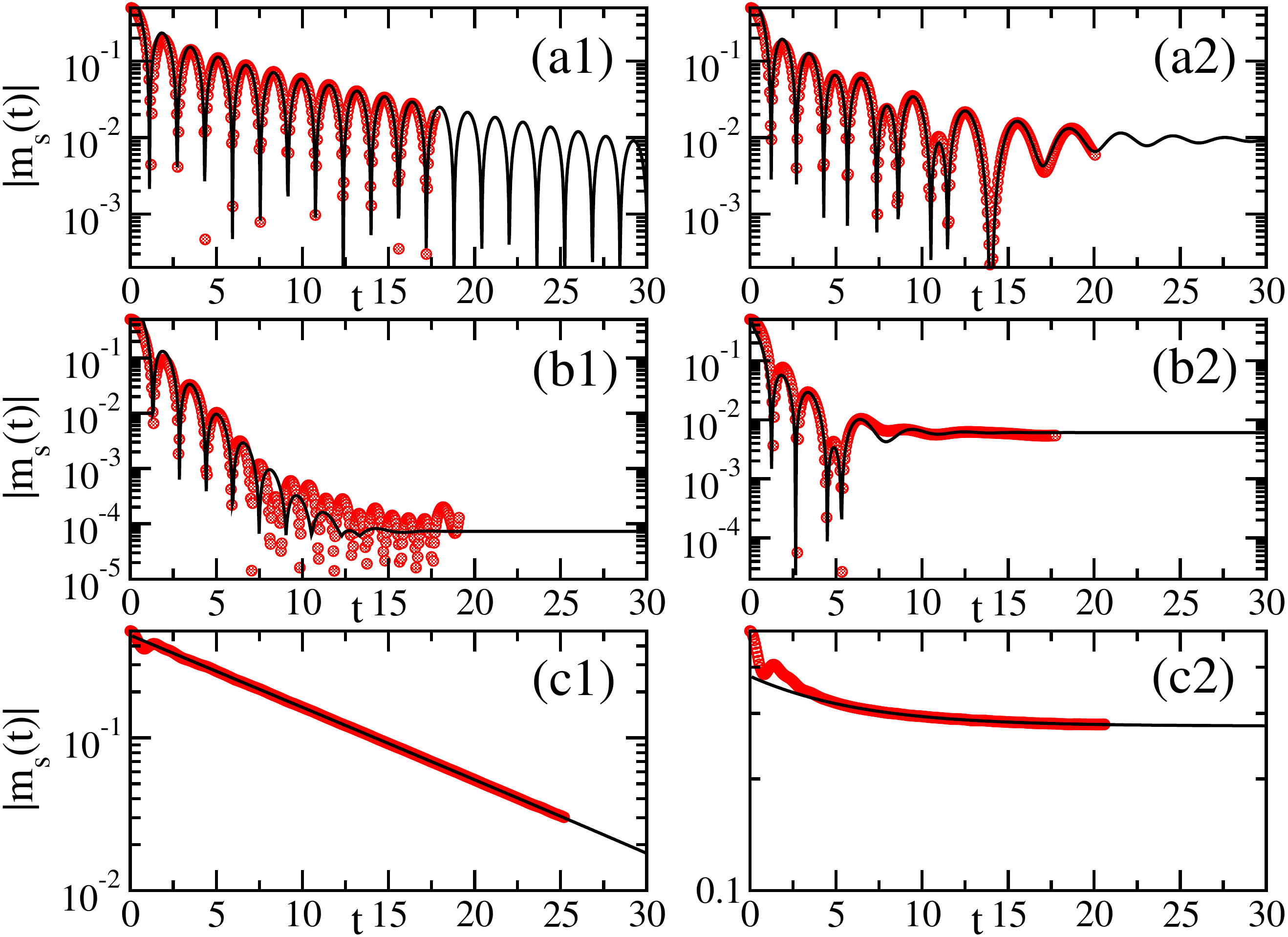}
 \caption{Decay of the order parameter $m_s(t)$ from LCRG for (a)
 $\Delta=0.25$, (b) $\Delta=1$, and (c) $\Delta=4$. Left column
 $D=0.02$ and right column $D=0.3$. Error bars are smaller than the
 symbol size. The lines are fits with absolute statistical errors for
 $m_s(\infty)$ of the order of $10^{-3}-10^{-4}$. The relaxation time
 $\tau$ decreases with increasing disorder for $\Delta=0.25$ and
 $\Delta=4$ while it increases for $\Delta=1$.}
\label{Fig_ms_vs_t}
\end{figure}

\section{Decay of order parameter}
\label{SecII}

In the clean free fermion case ($D=\Delta=0$) the decay of the order
parameter is given by
$m_s(t)=\frac{1}{2}J_0(2t)\sim (4\pi t)^{-1/2}\cos(2t-\pi/4)$ with
$J_0$ being the Bessel function of the first kind and time measured in
units of $\hbar/J$. For interactions $0<\Delta<1$ it has been shown
that the asymptotic decay in the clean case is well described by the
free fermion asymptotics multiplied by an exponential decay
\cite{BarmettlerPunk,BarmettlerPunk2}. Turning on disorder introduces
barriers between thermalizing clusters with equal Zeeman field, see
inset of Fig.~\ref{Fig_phase_diag}(a). In the ergodic phase, a finite
thermalization time across such barriers $\tau\sim \e^{Nf(D,\Delta)}$
must exist where $N$ is the number of jumps of the Zeeman field within
the barrier and $f(D,\Delta)$ a function depending on disorder $D$ and
anisotropy $\Delta$. The probability that a particular site is part of
a barrier with $N$ jumps is given by $P(N)=N/2^{N+1}$. After time $t$
only clusters separated by barriers of size
$N\geq N_0 = f^{-1}(D,\Delta)\ln t$ will not have thermalized, and the
asymptotic decay in the ergodic phase follows
\begin{equation}
\label{asymp}
m_s(t)\sim\int_{N_0}^\infty \!\!\! P(N) dN \sim \int_{f^{-1}(D,\Delta)\ln t}^\infty \frac{N\,dN}{2^{N+1}}\sim t^{-\frac{\text{const}}{f(D,\Delta)}}
\end{equation}
up to logarithmic corrections. In the MBL phase, on the other hand,
the staggered magnetization will not decay completely,
$m_s(\infty)\equiv m_s(t\to\infty)\neq 0$. Combining the different
limiting cases, we fit the LCRG data for anisotropies $0<\Delta\leq
1.25$, disorder $0<D<1$, and times $t\geq 5$ by the functions
\begin{equation}
\label{fit1}
m_s(t) = A\frac{\cos(\omega t-\phi)\e^{-t/\tau}}{\sqrt{t}}+\left\{\begin{array}{l} Bt^{-\zeta}\\ m_s(\infty)\end{array}\right.
\end{equation}
with lifetime $\tau$ and exponent $\zeta$ of a power-law decay. We
perform fits using both fit functions and check for consistency, i.e.,
in the ergodic phase $m_s(\infty)\approx 0$ and in the MBL phase
$\zeta\approx 0$ with $B\approx m_s(\infty)$. As shown in
Fig.~\ref{Fig_ms_vs_t}, this leads to excellent fits which allow to
extract an estimate for $m_s(\infty)$ in the MBL phase, $\zeta$ in the
ergodic phase, as well as the lifetime of the oscillations $\tau$, see
Fig.~\ref{Fig_relax_time}(a,b,c).  In Appendix \ref{sec:fits} we show
that the fit parameters depend only weakly on the time window used for
the fit, cf.\ Fig.~\ref{Fig5_add1}.  In particular, the asymptotic
value of the order parameter $m_s(\infty)$ is very robust in all fits,
cf.\ Fig.~\ref{Fig5_add2}.  As in the clean case \cite{BarmettlerPunk,
  BarmettlerPunk2}, we cannot find any fitting function which
describes the data for small disorder in the regime
$1.25<\Delta\leq 3$ well. For $\Delta\geq 3$, on the other hand, we
find that the asymptotics is very well described by a pure
non-oscillating exponential decay
$m_s(t)\sim m_s(\infty)+A\e^{-t/\tau}$, see Fig.~\ref{Fig_ms_vs_t}(c).

Based on
the RG analysis of a minimal model, an exponential sensitivity of the
residual imbalance $m_s(\infty)\sim
m_0\exp\left[-(D-D_c)^{-\nu}\right]$ in the MBL phase ($D>D_c$) has
been predicted \cite{PotterVasseur}.
\begin{figure}
 \includegraphics*[width=1.0\columnwidth]{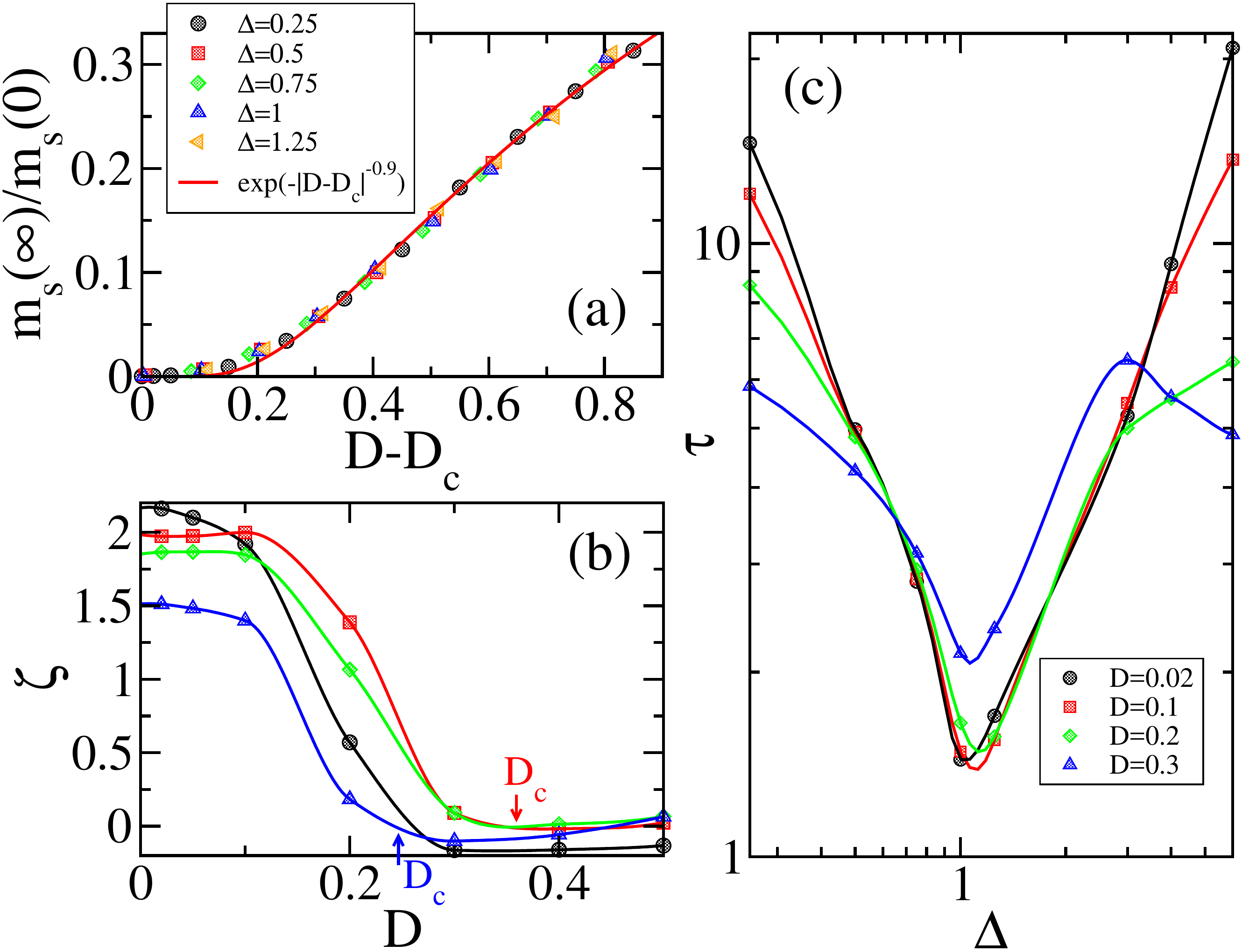}
 \caption{(a) Data collapse of the magnetization $m_s(\infty)$. (b)
 Power-law exponent $\zeta$ (same symbols as in (a)), and (c)
 relaxation times $\tau$ for small disorder $D$. Symbols: results from
 fits of the LCRG data; lines in (b) and (c) are guides to the eye.}
\label{Fig_relax_time}
\end{figure}
As shown in Fig.~\ref{Fig_relax_time}(a), we obtain an excellent data
collapse for different $\Delta$ with a critical exponent $\nu\sim 0.9$
using $m_0\in [0.255,0.282]$ and $D_c$ as fitting parameters. The
critical values $D_c(\Delta)$ obtained from the data collapse lead to
the phase boundary shown in Fig.~\ref{Fig_phase_diag}(a). We note that
$\nu\sim 0.9<2$ violates the Harris bound, see below. For comparison,
the power-law exponent $\zeta$ is shown in
Fig.~\ref{Fig_relax_time}(b): In a theory with a single length scale
$\xi$, one would expect that $\zeta\sim 1/z\sim 1/\xi\sim |D-D_c|^\nu$
where $z$ is the dynamical critical exponent
\cite{PotterVasseur}. However, the fits yield absolute statistical errors 
in the power-law exponent $\zeta$ between $0.05-0.2$ making it
impossible to extract a $\zeta(D)$ scaling close to $\zeta(D)\sim
0$. The $D_c$ values determined by $\zeta(D_c)=0$ nevertheless are
consistent with, although slightly larger than, the values based on
the data collapse for the magnetization. The relaxation time $\tau$,
on the other hand, can be extracted with statistical errors of less
than $2\%$ and is shown in Fig.~\ref{Fig_relax_time}(c). For very
small disorder we qualitatively find the same behavior as in the clean
case
\cite{BarmettlerPunk, BarmettlerPunk2}. The relaxation time decreases
approximately as $\tau\sim\lvert\ln\Delta\rvert$ for $\Delta<1$ and
increases proportional to $\tau\sim\Delta^2$ for $\Delta\geq 3$. For
$\Delta\ll 1$ and $\Delta> 3$ we find that the relaxation times
immediately decrease when disorder is added; in a region around
$\Delta\sim 1$, however, the relaxation times remain stable at first
before {\it increasing} at larger disorder strengths. When plotting
the value of the smallest disorder where $\tau$ deviates substantially
(by more than $10\%$) from the clean case for different anisotropies
$\Delta$, we find that this change in relaxation time does occur when
crossing from the ergodic to the MBL phase, see
Fig.~\ref{Fig_phase_diag}(a). 

Similarly to the phase diagram for the XXZ chain with box
disorder---obtained by ED in Ref.~\onlinecite{BarLev}---we observe
reentrant behavior: for fixed $D$ and small $\Delta$ in the MBL phase,
increasing interactions can first drive the system into the ergodic
phase before localization is again stabilized at large interactions.

\section{Entanglement growth}

\begin{figure}
 \includegraphics*[width=1.0\columnwidth]{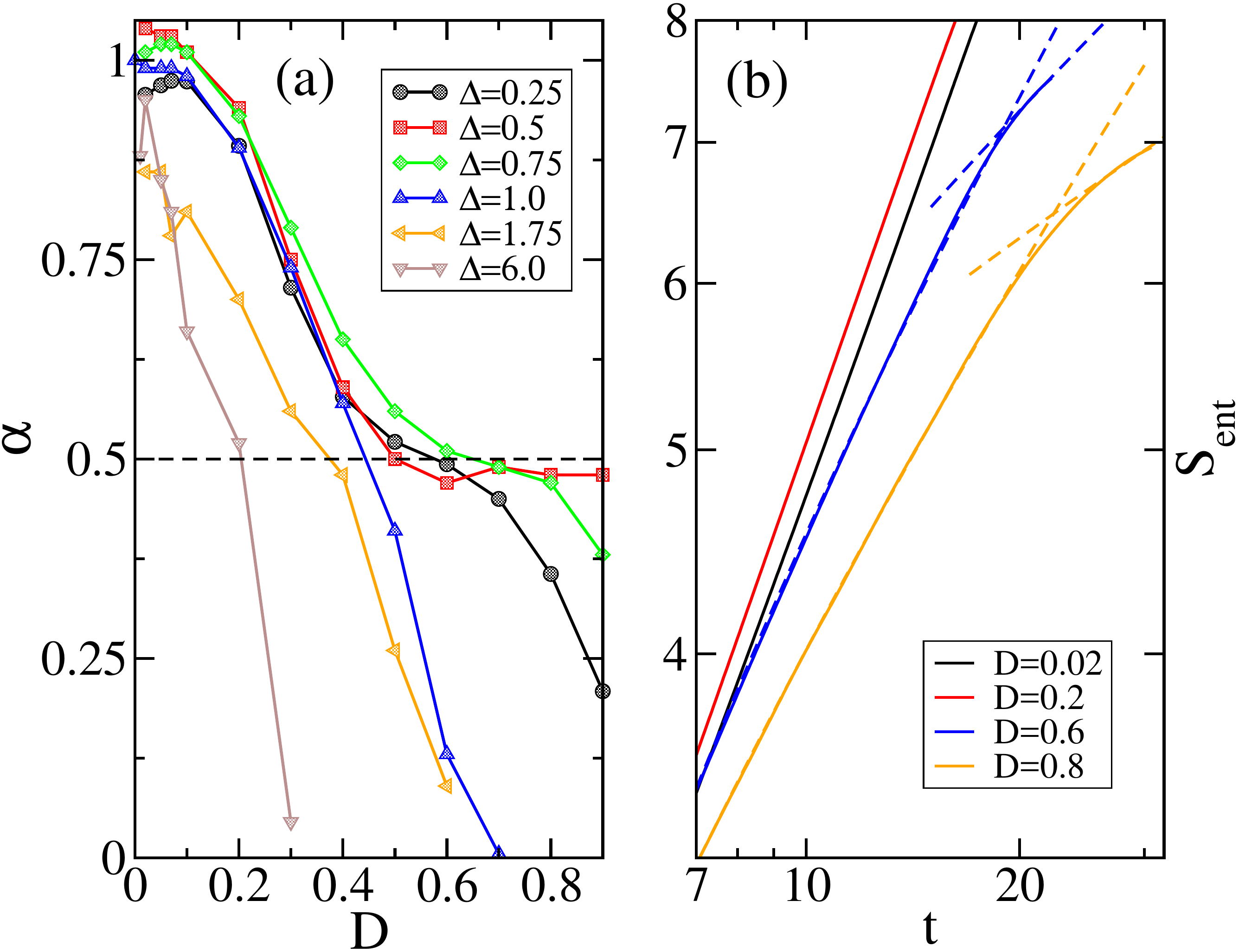}
 \caption{Entanglement entropy: (a) Exponent
   $S_{\textrm{ent}}\sim t^\alpha$ for different anisotropies $\Delta$
   at the longest times accessible by LCRG. (b) $S_{\textrm{ent}}(t)$
   for $\Delta=1$ and different disorder strengths $D$ on a log-log
   scale. The dashed lines are power-law fits.}
\label{Fig_entanglement}
\end{figure}
To investigate the properties of the phase transition in more detail,
we now turn to an analysis of the entanglement entropy
$S_{\textrm{ent}}(t)$. Using the same type of argument as for the
decay of the order parameter, a power law
$S_{\textrm{ent}}\sim t^\alpha$ is expected in the ergodic
phase; a power law in the entanglement entropy is also found
in the critical Harper model \cite{RooszIgloi}. If the RG theories of
Refs.~\onlinecite{VoskHuse, PotterVasseur} do describe the transition
correctly then $\alpha=1/z\sim 1/\xi$ holds. On the MBL side, on the
other hand, we have shown previously in
Ref.~\onlinecite{AndraschkoEnssSirker} that $S_{\textrm{ent}}\sim\ln
t$ as is predicted on general grounds.\cite{VoskAltman, SerbynPapicA,
  SerbynPapicB, NanduriKim, HuseNandkishore} For $\Delta<1$ we fit the
LCRG data for $t>7.5$ to a power law $S_{\textrm{ent}}=c_1+c_2
t^\alpha$ and obtain excellent fits with statistical errors of less
than $5\%$. Furthermore, the exponent $\alpha$, shown in
Fig.~\ref{Fig_entanglement}(a), is only weakly affected by a
modification of the fit interval provided that the behavior for small
times is excluded. For $\Delta \geq 1$ and intermediate disorder, on
the other hand, we find two different regimes: a power law increase at
intermediate times $5<t\lesssim 20$ followed by a much slower increase
for $t\gtrsim 20$, see Fig.~\ref{Fig_entanglement}(b). Because of the
limited time range available, it is not clear if the latter regime
corresponds to a power-law increase with a smaller exponent or signals
a crossover to logarithmic scaling.

Overall, we find a sub-ballistic spreading including, surprisingly, an
extended region of disorder strengths for small $\Delta$ where the
entanglement spreads diffusively, $\alpha=1/2$ and
$S_{\textrm{ent}}\sim \sqrt{t}$. Remarkably, for the parameters
$\Delta=0.5$, $D=0.5,\dotsc,0.9$ where the entanglement spreads
diffusively the system seems to be already deep in the MBL phase
according to the phase diagram Fig.~\ref{Fig_phase_diag}(a); the order
parameter shown in Fig.~\ref{Fig5_add2} signals localization at least
up to times $t\sim 25$. Physically, this intermediate diffusive regime
might be explained by the existence of many relatively narrow barriers
between thermalizing segments which lead to diffusion while rare wide
barriers lead to an exponential enhancement of the entanglement time
and finally prevent the system from fully thermalizing. Our findings
might therefore possibly indicate that the transition is not described
by a single length scale. In this case the scaling hypothesis is
violated and a Harris criterion does not apply. The region with $z=2$
corresponds to classical diffusion implying, in particular, that the
spin-spin autocorrelation function decays as $C_{zz}(t)\sim
t^{-\beta}$ with $\beta=1/z=1/2$. Note that even in the clean case the
presence or absence of diffusion at infinite temperatures in
$C_{zz}(t)$ is an open and controversially discussed topic, with
numerical results for short and intermediate times showing a power-law
decay with an exponent depending on the fit interval
\cite{FabriciusMcCoy, SirkerDiff}. At small finite temperatures, on
the other hand, $C_{zz}\sim 1/\sqrt{t}$ has recently been established
by field theoretical methods in the TDL and confirmed by numerical
data
\cite{KarraschPereiraSirker}.

\section{Comparison with ED}
\label{SecIV}
\begin{figure}
 \includegraphics*[width=1.0\columnwidth]{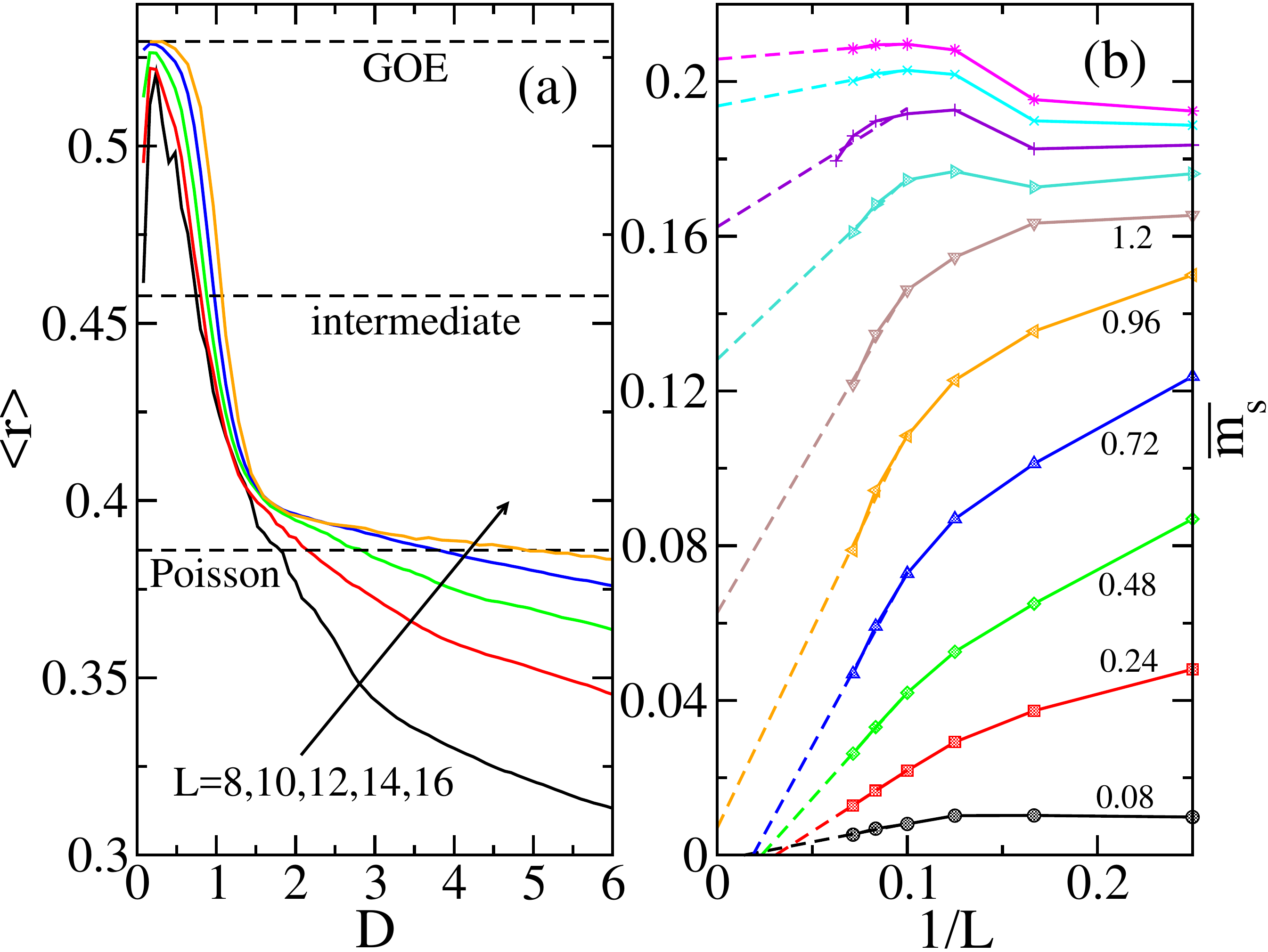}
 \caption{(a) $\Delta=1$: $\langle r\rangle$ values for open chains of
 length $L$. (b) Time averaged magnetizations $\overline{m_s}$ for disorder
 $D=0.08,0.24,0.48,0.72,\dotsc,2.16$ (from bottom to top). The dashed
 lines are linear extrapolations in $1/L$ for $L\geq 10$.}
\label{Fig_ED}
\end{figure}
While the LCRG data for infinite systems support a consistent
interpretation of the MBL transition, it is instructive to compare to
exact diagonalization results for finite systems. Two commonly used
methods to establish the phase diagram of the disordered model
\eqref{Ham} are calculating the level statistics and studying the time
average of an order parameter.

To obtain the level statistics we define
$r_n=\min(\delta_n,\delta_{n-1})/\max(\delta_n,\delta_{n-1})$ with
$\delta_n=E_{n+1}-E_n$ the difference between adjacent energy
eigenvalues. At the integrable point $D=0$ and also in the MBL phase
where additional local conserved charges exist we expect Poisson
statistics $P(r)=2/(1+r)^2$ with an average value
$\langle r\rangle \approx 0.386$, while Wigner-Dyson statistics with
$\langle r\rangle \approx 0.529$ is expected in the ergodic phase for
$D\neq 0$ \cite{oganesyan2007,PalHuse}, see Appendix \ref{sec:ed} for
details.  In Fig.~\ref{Fig_ED}(a), results for model \eqref{Ham} with
binary disorder, $\Delta=1$, and system sizes $L=8-16$ are shown where
the disorder averages are {\it exact} for $L\leq 14$ while $4000$
inequivalent samples have been used for $L=16$. Contrary to the box
disorder case \cite{PalHuse} we do not find a point where
$\langle r\rangle(L)$ appears to be close to stationary which has been
interpreted as being indicative of the critical point in the
thermodynamic limit. Note, however, that even in the box disorder case
a `drifting' of the crossing points of curves with different $L$ has
been observed for increasing system size.

An alternative criterion to estimate the phase boundary is to fix
$D_c(L)$ as the disorder value where $\langle r\rangle(L)$ crosses the
intermediate value $\langle r\rangle=0.4575$ between Wigner-Dyson and
Poisson statistics. If there is a sharp transition between an ergodic
and an MBL phase in the thermodynamic limit then $D_c(L)$ will
converge to the critical value in the limit $L\to\infty$. However,
even using this alternative criterion the problem persists that no
clear scaling with $L$ is obtained for the limited system sizes
available. For $L=16$, $\langle r\rangle$ takes an intermediate value
between Wigner-Dyson and Poisson statistics around disorder $D\sim 1$
(see Fig.~\ref{Fig_r_values} in the appendix), which is an order of
magnitude larger than the $D_c$ value for $\Delta=1$ established above
for the infinite chain.

With increasing anisotropy $\Delta$ the system approaches the Ising
limit where each local $S^z_j$ becomes approximately conserved. For
small disorder it would then require very large systems to see level
repulsion and Wigner-Dyson statistics.  We therefore consider only
anisotropies $\Delta\lesssim 2$ using ED, see
Fig.~\ref{Fig_phase_diag}(b).
\begin{figure}
  \includegraphics*[width=1.0\columnwidth]{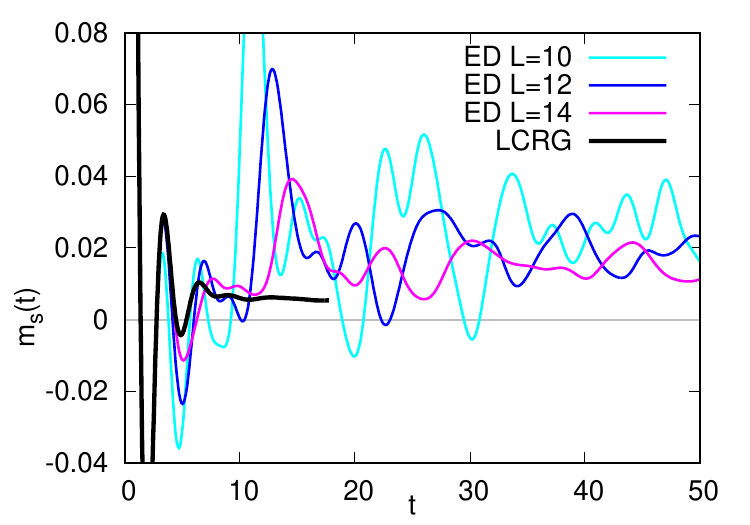}
  \caption{Time evolution of the staggered magnetization $m_s(t)$ at
    the isotropic point $\Delta=1$ for small disorder $D=0.3$ from ED
    for $L\leq14$ with periodic boundary conditions (PBC) and LCRG
    ($L=\infty$).  Already for short times $Jt\sim4$ the time
    evolution differs visibly due to finite-size effects in ED.}
  \label{Fig_timeevol03}
\end{figure}

A naive linear extrapolation in $1/L$ of the time averaged
magnetizations $\overline{m_s}$ also yields a critical $D_c\sim 1$ for
$\Delta=1$, see Fig.~\ref{Fig_ED}(b). LCRG, on the other hand, shows
quite clearly that $D=1$ is already deep inside the MBL phase (see
Fig.~\ref{Fig_ms_vs_t}(b2)). Using both $\langle r\rangle$ and
$\overline{m_s}$ to extract a phase boundary shows that the ED results
can lead to a significantly larger extent of the ergodic phase for all
$\Delta$, see Fig.~\ref{Fig_phase_diag}. This might not be completely
unexpected because any system with length $L$ much smaller than the
localization length $\xi_\text{loc}$ will look ergodic.

The difference between infinite and finite systems is exemplified
clearly in the time evolution of the order parameter $m_s(t)$ for
$\Delta=1$, $D=0.3$ shown in Fig.~\ref{Fig_timeevol03}.  According to
the finite-size scaling of the time averaged $\overline{m_s}$ obtained
by ED, this is far in the ergodic phase (cf.\ Figs.~\ref{Fig_ED}(b)
and \ref{Fig_Magscaling}).  However, we observe that ED for $L\leq14$
and LCRG for $L=\infty$ only agree up to $t\lesssim4$.  LCRG, instead,
shows that $m_s(t)$ saturates for $t\gtrsim10$ at least up to
$t\sim16$, which corresponds to an effective system size
$L\sim2vt\sim64$.  Since the LCRG data for the infinite chain test the
dynamics at length scales which are a factor $4-6$ larger than the
length scales reached in ED, the most plausible explanation for this
discrepancy appears to be that the scaling of $D_c(L)$ is
non-monotonic.  In order to check this tentative explanation, one
would need to diagonalize much larger systems.

\section{Conclusions}
Using time-dependent density matrix renormalization group calculations
we have established the phase diagram of the XXZ spin-$1/2$ chain with
binary disorder in the TDL. For weak disorder in the ergodic phase
we are able to test the dynamics on length scales of the order of
$100$ lattice sites which is significantly larger than the lengths
which can be studied in exact diagonalization. Our results generalize
previous studies of the decay of N\'eel order (imbalance), $m_s(t)$,
from clean to disordered systems which is highly relevant to interpret
recent
\cite{SchneiderBloch1, SmithMonroe, SchneiderBloch2} and future cold
atomic gas experiments. We find that $m_s(\infty)$ in the MBL phase
shows an exponential sensitivity on disorder with a critical exponent
near the ergodic-MBL phase transition of $\nu\sim 0.9$. For the
entanglement entropy $S_{\textrm{ent}}(t)$ we find a power-law growth
at intermediate times with an exponent which varies continuously as a
function of disorder. For small $\Delta$ we find, in particular, a
diffusive growth of entanglement $S_\text{ent}\sim\sqrt t$ at
intermediate times in the MBL phase near the transition while
$S_{\textrm{ent}}\sim\ln t$ is expected at long times. This
intermediate time behavior might indicate a second relevant length
scale in the problem. In this case the scaling hypothesis is violated
and a Harris bound $\nu\geq 2$ does not apply.

\acknowledgments
J.S. acknowledges support by the Natural Sciences and Engineering
Research Council (NSERC, Canada) and by the Deutsche
Forschungsgemeinschaft (DFG) via Research Unit FOR 2316.  We are
grateful for the computing resources and support provided by Compute
Canada and Westgrid.

\appendix

\section*{Appendix}
In the Appendix we provide technical details regarding the fitting of
the LCRG data and the exact diagonalizations. Furthermore, we present
spectra of time averaged magnetizations for individual disorder
realizations which show qualitative differences in the ergodic and
deep in the MBL phase and might be a useful tool for experimental
analysis.

\subsection{Fits of the order parameter}
\label{sec:fits}
Using the LCRG algorithm we have obtained data for the decay of
$m_s(t)$ directly in the thermodynamic limit. To analyze these data we
have used the two fit functions given in Eq.~\eqref{fit1}. Here we
want to show that these fits are quite stable with regard to the time
window chosen provided that one excludes the initial fast decay which
is not well described by the fit functions. We always start with the
data for the smallest disorder, $D=0.02$, using the values for the
free fermion case without disorder $A=1/\sqrt{4\pi}$, $\omega=2$,
$\phi=\pi/4$, $\tau=\infty$, and $B=\zeta=m_s(\infty)=0$ as initial
guess. The fitting parameters obtained from the converged least square
fit are then used as initial parameters for the fit of the data set
with the next larger disorder.

\begin{figure}
  \includegraphics*[width=1.0\columnwidth]{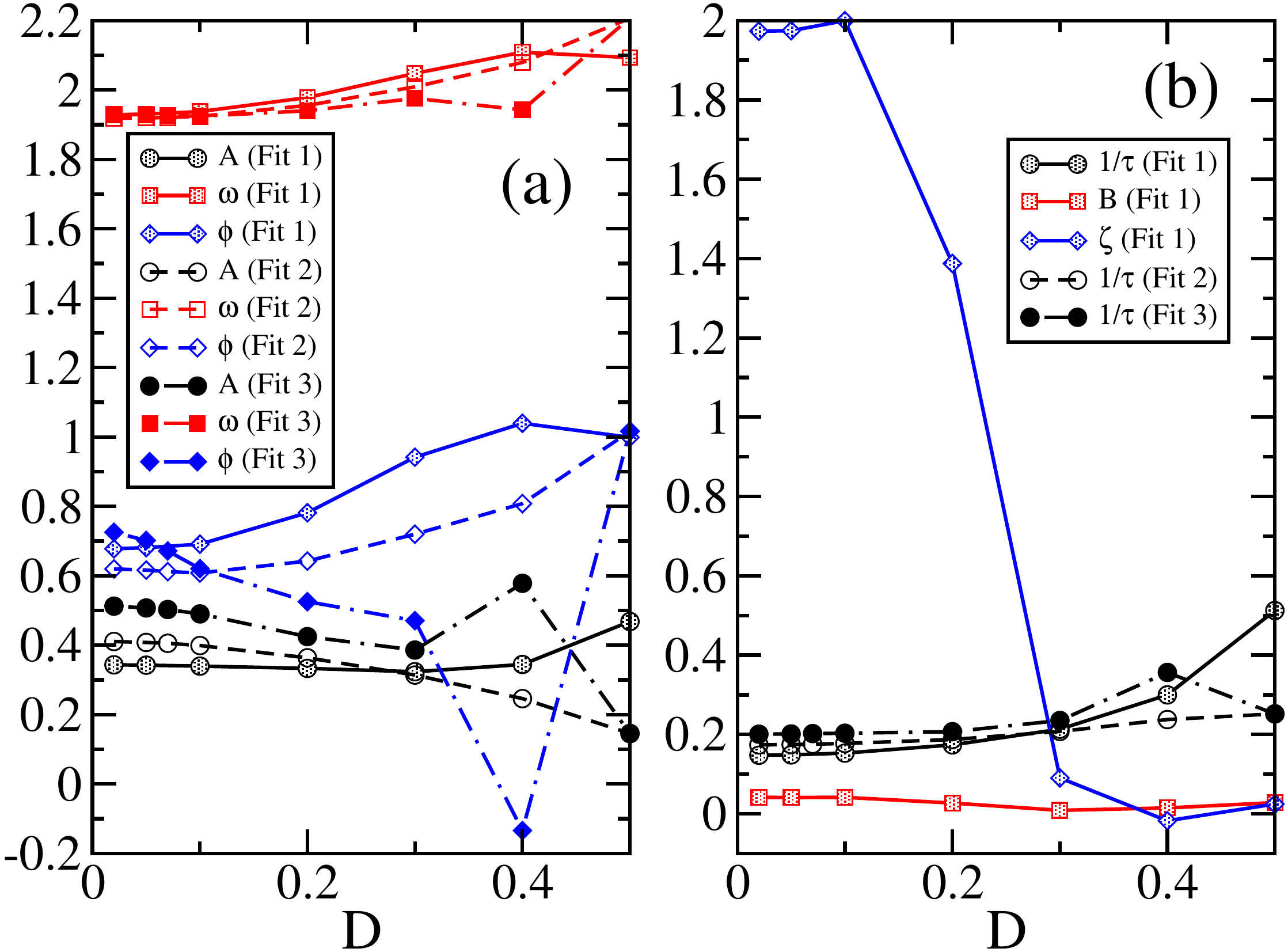}
  \caption{Fit parameters for $\Delta=0.5$ using the fit function
    Eq.~\eqref{fit1}. Fit 1: $t\geq 1$ with power law $Bt^{-\zeta}$,
    fit 2: $t\geq 3$ with $m_s(\infty)$, and fit 3: $t\geq 5$ with
    $m_s(\infty)$. The values obtained for $m_s(\infty)$ are shown in
    Fig.~\ref{Fig5_add2}(b).}
  \label{Fig5_add1}
\end{figure}

As an example, we show here different fits and additional data for the
particularly interesting case $\Delta=0.5$. In Fig.~\ref{Fig5_add1} we
compare the parameters of three different fits.
For $D\leq 0.2$ all three fits give parameters which are quite close
to each other and which change smoothly as a function of disorder. For
larger disorder values most parameters remain stable except for the
phase shift $\phi$. We want to stress again that the fit functions are
based on the asymptotics for the clean free fermion case and are not
expected to yield good fits for large disorder and/or interaction
strengths. For $\Delta=0.5$ we can obtain reasonable fits up to
disorder $D=0.5$. 

\begin{figure}
  \includegraphics*[width=1.0\columnwidth]{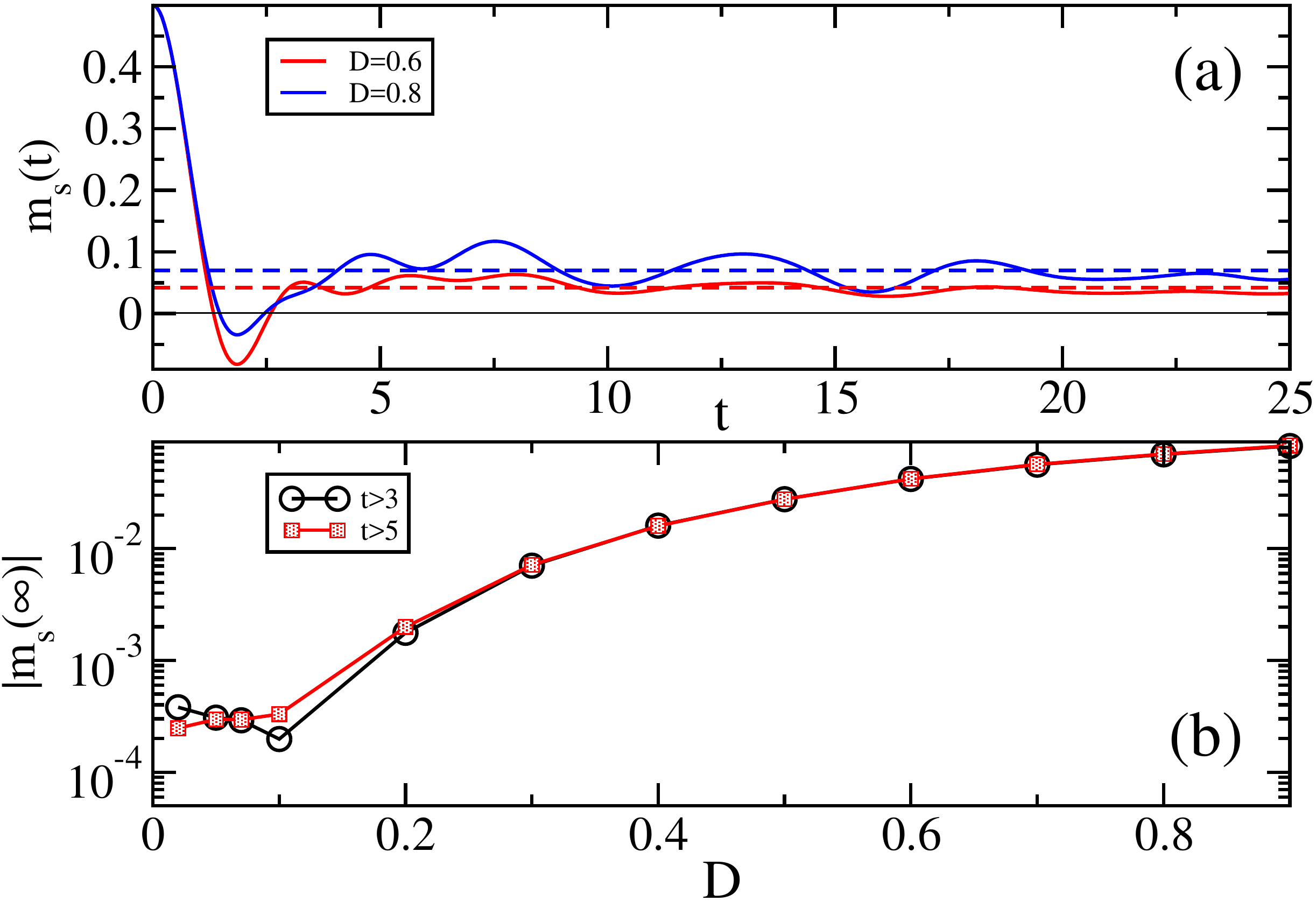} 
\caption{(a)
  $m_s(t)$ for $\Delta=0.5$ and $D=0.6,0.8$. The dashed lines denote
  the time average for $t\in [5,t_{\textrm{max}}]$. (b) $m_s(\infty)$
  extracted from the fits (see Fig.~\ref{Fig5_add1} for the other
  fitting parameters) for $D\leq 0.5$ and from a time average for
  $D>0.5$.}  \label{Fig5_add2}
\end{figure}

For $D\leq 0.5$ we extract the remaining magnetization at infinite
times, $m_s(\infty)$, from the fits while for $D>0.5$ we simply take the
average of $m_s(t)$ in the specified time window. The results are
shown in Fig.~\ref{Fig5_add2}(b) and are almost independent of the
time window.
We have found that there appears to be a diffusive entanglement
spreading for $\Delta=0.5$ and $D\in [0.5,0.9]$, see
Fig.~\ref{Fig_entanglement}. From the magnetization data it appears,
however, that for these disorder strengths we are already in the MBL
phase, see Fig.~\ref{Fig5_add2}(a). A possible interpretation is that
the diffusive entanglement spreading only holds at intermediate times
while a crossover to the expected logarithmic scaling will happen at
larger times, inaccessible to our numerical calculations.

\subsection{Exact Diagonalization (ED)}
\label{sec:ed}
We use ED for small XXZ chains of length $L$ to study the level
statistics of the disordered Hamiltonian ($r$ values) as well as the
time evolution of observables such as the staggered magnetization,
$m_s(t)$.

For each disorder configuration the Hamiltonian \eqref{Ham} conserves
the total spin quantum number $S^z = \sum_i s_i^z$.  Here, we consider
chains with no average magnetization, $S^z=0$, for even $L$.  The
energy spectrum $E_n$ determines the level statistics, while all
eigenvectors are needed for expectation values such as the staggered
magnetization.  The computation is repeated for different disorder
realizations and the results averaged.  In particular, in the case of
binary disorder, we explicitly average over all
$\mathcal N_\text{dis} = 2^L$ possible disorder configurations.  By
symmetry, the configurations with flipped disorder $D_i \mapsto -D_i$
or with left-right mirrored disorder $D_i \mapsto D_{L+1-i}$ yield
equivalent results.  This reduces the number of inequivalent
configurations to $\sim 2^L/4$ in the case of open boundary conditions
(OBC).  For periodic boundary conditions (PBC), the shift symmetry
$D_i \mapsto D_{i+1}$ leads to a further reduction to $\sim 2^L/4L$.
For instance for $L=16$, the dimension of the $S^z=0$ Hilbert space is
$12870$, and there are $16512$ (OBC) and $1162$ (PBC) unique disorder
configurations, resp.  For $L\leq14$ and $L=16$ (PBC) we typically
perform complete disorder averages; for $L=16$ (OBC) we sample $4000$
inequivalent random configurations.  This is in contrast with the LCRG
algorithm which works in the much larger Hilbert space of spins and
ancillas and produces the complete disorder average in a single run
\cite{AndraschkoEnssSirker}.

\subsubsection{ED Level statistics}
For each disorder configuration we define the level spacing $\delta_n
= E_{n+1}-E_n$ between adjacent energy eigenvalues $E_n$. In order to
normalize the energy scale we consider the ratios $r_n =
\min(\delta_n,\delta_{n-1})/\max(\delta_n,\delta_{n-1})$ which lie
between $0$ and $1$.  The level distribution $P(r)$ is then averaged
over all binary disorder configurations. In the presence of an
extensive set of local conserved charges the level spacing $\delta_n$
is Poisson distributed with $P(r)=2/(1+r)^2$ and average value
$\langle r\rangle_\text{Poisson}
\approx 0.386$.  In the ergodic phase, instead, a Wigner-Dyson
distribution (GOE) of $\delta_n$ favors larger ratios with $\langle
r\rangle_\text{GOE} \approx 0.529$.  

\begin{figure}
  \includegraphics*[width=1.0\columnwidth]{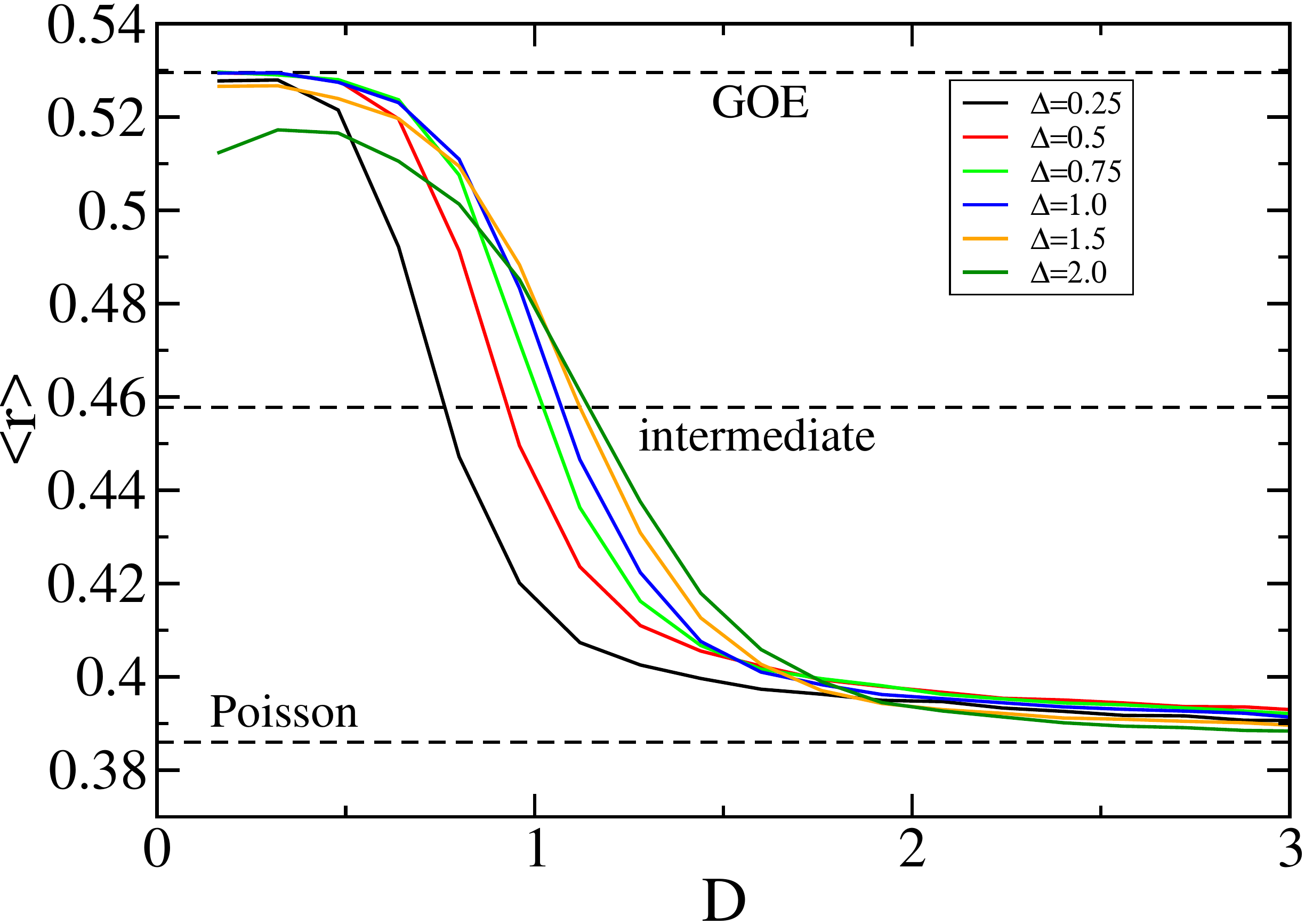}
  \caption{$\langle r \rangle$ values for open XXZ chains for
  different anisotropies $\Delta$ with $L=16$. The critical disorder
  strength where $\langle r\rangle$ crosses over from GOE to Poisson
  is used as criterion for the ergodic-MBL phase transition.}
  \label{Fig_r_values}
\end{figure}

Level spectra for open boundary conditions show less degeneracies as
compared to those for periodic boundary conditions and are therefore
better suited to determine the phase boundary. In
Fig.~\ref{Fig_r_values} we show the $\langle r \rangle$ values for
different anisotropies $\Delta$ as a function of disorder $D$ at fixed
$L=16$. The points where these curves cross the intermediate $\langle
r\rangle_{\textrm{crit}}\approx 0.4575$ determine the phase boundary
shown in Fig.~1(b) in the main text.

\subsubsection{ED magnetization}
\paragraph{Time evolution.}
The time evolution of the staggered magnetization from an initial
N\'eel state $\lvert\Psi(0)\rangle$ is computed as the disorder
average of the quantum evolution
\begin{equation*}
  m_s(t) = \langle \langle\Psi_0 \rvert e^{iHt} \Hat m_s e^{-iHt}
  \lvert \Psi_0 \rangle \rangle_\text{dis}.
\end{equation*}
In the eigenbasis $\lvert\phi_i\rangle$ for each disorder
configuration, one can write
\begin{equation*}
  m_s(t) = \Bigl\langle \sum_{ij} e^{i(E_j-E_i)t}
    \langle\Psi(0) \vert \phi_j\rangle
    \langle\phi_j\rvert \Hat m_s \lvert\phi_i\rangle
    \langle\phi_i \vert \Psi(0)\rangle \Bigr\rangle_\text{dis}.
\end{equation*}
Fig.~\ref{Fig_timeevol03} shows the time evolution of the staggered
magnetization for a Heisenberg chain $\Delta=1$ with small disorder
$D=0.3$.  The LCRG results are exact for an infinite system $L=\infty$
and extend to finite times $Jt\sim 18$. They provide strong evidence
that $m_s(\infty)\neq 0$ and that the system is therefore in the MBL
phase in accordance with the phase diagram
Fig.~\ref{Fig_phase_diag}(a) in the main text. The ED time evolution
for $L\leq14$ can be computed for arbitrarily long times but deviates
from the LCRG $L=\infty$ result already for short times $Jt\sim4$ due
to finite-size effects.  The localization length just beyond the MBL
transition is much larger than any system size accessible by ED so
that ED can only capture the short-time dynamics correctly, making an
extrapolation of time averaged data to lengths
$L\gg\xi_{\textrm{loc}}$ impossible.

\begin{figure}
  \includegraphics*[width=1.0\columnwidth]{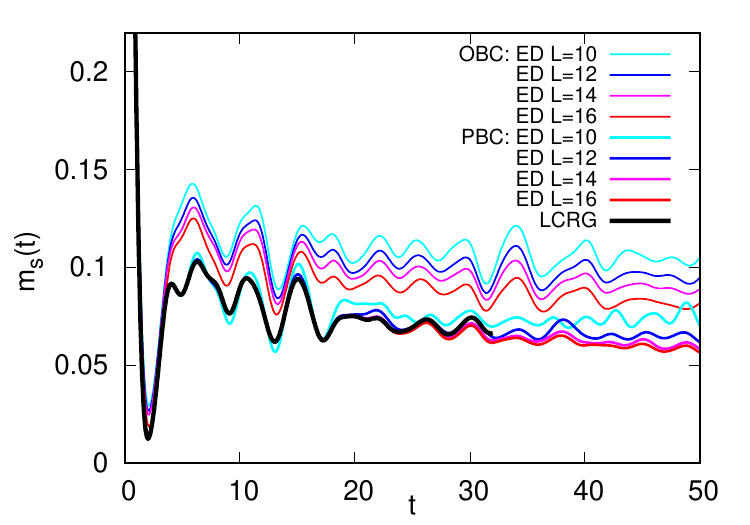}
  \caption{Time evolution of the staggered magnetization $m_s(t)$ at
  the isotropic point $\Delta=1$ for disorder $D=0.9$ from ED
  for $L\leq16$ with open (OBC) and periodic boundary conditions (PBC)
  and LCRG ($L=\infty$).  While OBC results converge slowly to the
  $L=\infty$ limit, the PBC results are quite accurate up to
  $Jt\sim20$. Disorder averages are exact except for $L=16$ with OBC
  where 950 samples have been used.} \label{Fig_timeevol09}
\end{figure}
For larger disorder $D=0.9$ shown in Fig.~\ref{Fig_timeevol09}, the ED
results for periodic boundary conditions are much closer to the LCRG
$L=\infty$ result and differ visibly only for $Jt\gtrsim20$ ($L=16$
PBC).  This is likely due to the proliferation of small localized
clusters which are well captured by ED and which dominate the dynamics
well inside the MBL phase.  In contrast, the ED time evolution for
open boundary conditions (upper set of curves) is far from the
$L=\infty$ result even for $L=16$ (OBC) and converges only slowly with
increasing $L$.

\begin{figure}
  \includegraphics*[width=1.0\columnwidth]{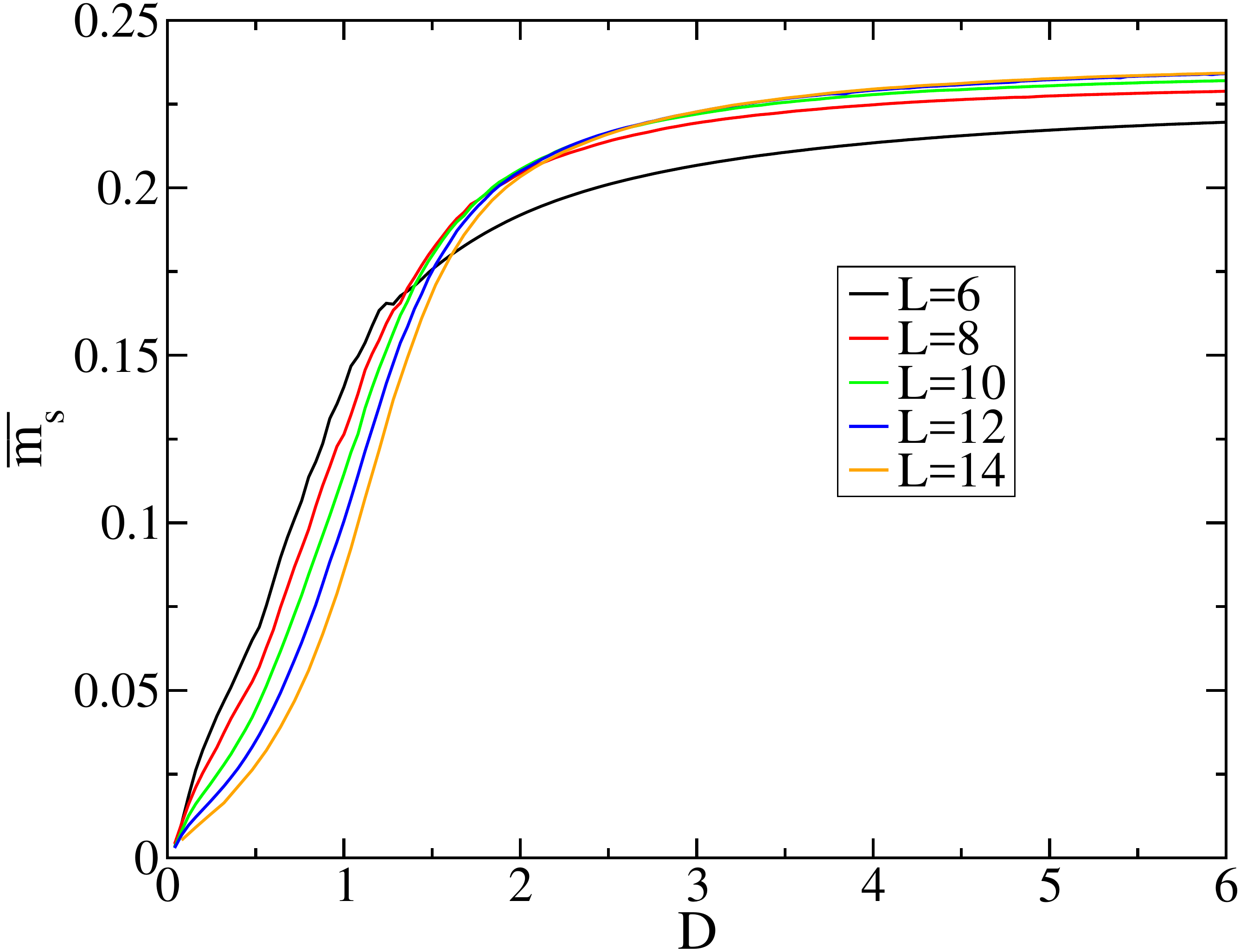}
  \caption{Time averaged magnetizations at the isotropic point
    $\Delta=1$ as a function of disorder $D$ for different lengths $L$
    (ED with OBC).}
  \label{Fig_Mag_Delta1p0}
\end{figure}
\begin{figure}[t]
  \includegraphics*[width=1.0\columnwidth]{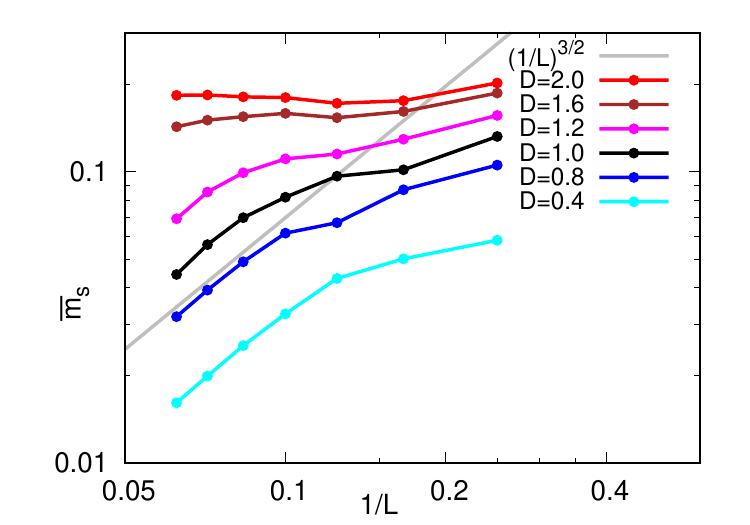}
  \caption{Time averaged magnetizations at the isotropic point
    $\Delta=1$ as a function of system size $1/L$ for different
    disorder $D$ (ED with PBC).  The finite-size scaling is compatible
    with $\overline{m_s}(L) \sim L^{-3/2}$.}
  \label{Fig_Magscaling}
\end{figure}
\paragraph{Time averaged magnetization.}
At long times, the staggered magnetization oscillates around the
average magnetization
\begin{align}
  \overline{m_s}
  & = \lim_{T\to\infty} \frac 1T \int_0^T dt\, m_s(t) \notag\\
  \label{eq:avmagn}
  & = \left\langle \sum_{ij} \delta_{E_i,E_j} \langle\Psi_0 \vert \phi_j\rangle
    \langle\phi_j\rvert \Hat m_s \lvert\phi_i\rangle
    \langle\phi_i \vert \Psi_0\rangle \right\rangle_\text{dis}.
\end{align}
The contributions with unequal energy dephase and do not contribute to
the time average, such that only the energy diagonal terms remain.
Note that matrix elements of $\Hat m_s$ between different degenerate
states vanish, such that the sum in \eqref{eq:avmagn} reduces to a
single sum with $i=j$.

The dependence of the average magnetization $\overline{m_s}$ on
disorder $D$ and system size $L$ is shown in
Fig.~\ref{Fig_Mag_Delta1p0}. A nonzero value $\overline{m_s}>0$ is
obtained for any disordered system with $L<\infty$.  Note that
$\overline{m_s}(L)$ decreases with increasing $L$ for small $D$, while
it increases for larger $D$.  While a crossing point exists, it does
not agree with the phase transition point found by a $1/L$ finite-size
scaling analysis in Fig.~\ref{Fig_ED}(b) nor with the phase boundary
obtained in LCRG for the infinite chain.

\begin{figure}
  \includegraphics*[width=1.0\columnwidth]{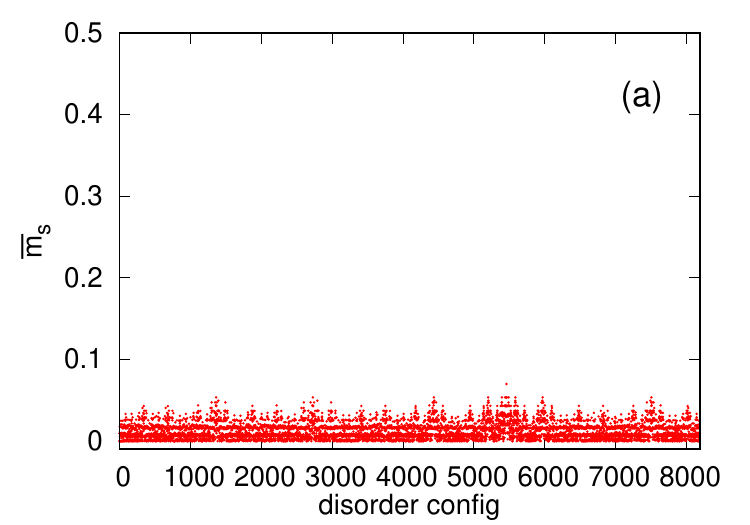} % was width=0.49
  \includegraphics*[width=1.0\columnwidth]{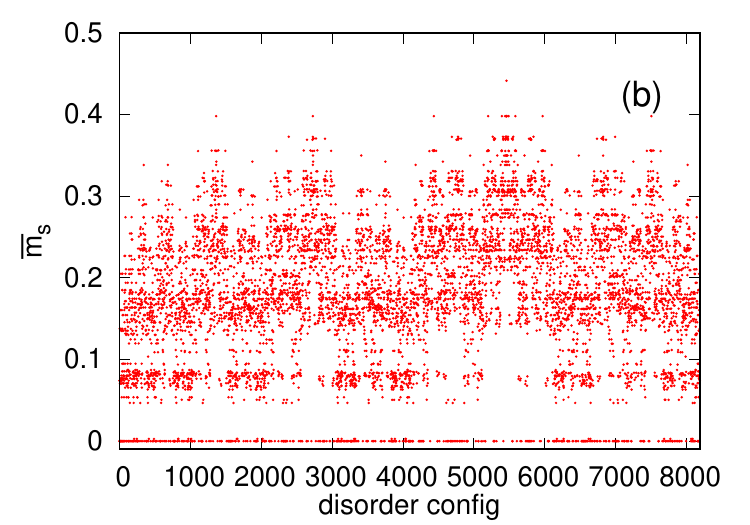}	
  \caption{$\overline{m_s}$ as a function of disorder configuration
  for a chain with length $L=14$, $\Delta=1$ (ED with PBC) where (a)
  $D=0.2$, and (b) $D=2.0$.}  
\label{Fig_spectra}
\end{figure}
A finite-size scaling analysis for the average magnetization
$\overline{m_s}(L)$ from ED with periodic boundary conditions (PBC) is
shown in Fig.~\ref{Fig_Magscaling} for the isotropic point.  In the
ergodic phase we expect the magnetization to decay for an infinite
system as $m_s(t)\sim t^{-1/z}$ where $z$ is the critical
exponent. Since the total magnetization is conserved, $\sum_j
S^z_j=\mbox{const}$, we expect that spin transport occurs as a random
walk similar to energy transport leading to a scaling $t(L)\sim
L^{z+1}$.\cite{VoskHuse,PotterVasseur} This scaling argument would
suggest that $\overline{m_s}(L) \sim L^{-(1+1/z)}$. For $D\lesssim
1.2$ we find that the scaling of the magnetization in
Fig.~\ref{Fig_Magscaling} appears to follow a power law with
exponent $1+1/z=3/2$, or $z=2$. This seems to further support our
findings from the analysis of the entanglement entropy for infinite
chains presented in the main text that the dynamics at intermediate
times (intermediate lengths) in the MBL phase close to the transition
is diffusive. For larger disorder the average magnetization saturates
to a finite value.  The apparent position of the MBL phase transition
with PBC is consistent with, but slightly larger than, the phase boundary
obtained for open boundary conditions as shown in Fig.~1(b) in the
main text.

\paragraph{Magnetization spectra.}
Using ED we can calculate a time averaged magnetization,
Eq.~\eqref{eq:avmagn}, for each disorder configuration. In
Fig.~\ref{Fig_spectra} the $\overline{m_s}$ values as a function of
the disorder configuration are exemplarily shown for $\Delta=1$ and
$D=0.2,2.0$.
The two magnetization spectra are qualitatively very different. While
the spectrum shown in Fig.\ref{Fig_spectra}(b) for large disorder
$D=2.0$ (deep inside the MBL phase) shows a gap, there is no gap for
$D=0.2$ (near the phase transition) visible, see
Fig.\ref{Fig_spectra}(a). For fixed $L=14$ we find that the gap for
$D=1.6$ is about a factor $10$ larger than the gap for $D=1.2$. This
provides an estimate for the phase transition which is consistent with
the estimate based on the level spectra for the same system size.

\end{document}